\documentclass[notitlepage,twocolumn,letterpaper,natbib,aps,prd,amsmath,amsfonts,nofootinbib,preprintnumbers,superscriptaddress,secnumarabic]{revtex4-1}
\pdfoutput=1
\usepackage{amssymb,amsmath,latexsym,mathrsfs}
\usepackage{url}
\usepackage{enumitem}
\usepackage{graphicx}
\usepackage{booktabs}
\usepackage[usenames,dvipsnames]{color}
\usepackage[breaklinks,colorlinks,urlcolor=magenta,citecolor=magenta,linkcolor=magenta]{hyperref}
\usepackage{multirow}
\usepackage{float}
\usepackage{cases}
\usepackage{blindtext}
\usepackage{pifont}
\usepackage{hhline}

\usepackage{xcolor}
\definecolor{linkcolor}{HTML}{FF6461}
\definecolor{citecolor}{rgb}{1.0, 0.5, 0.0}
\hypersetup{
  linkcolor  = linkcolor,
  citecolor  = linkcolor,
  urlcolor   = linkcolor,
  colorlinks = true
}

\definecolor{lcdmcol}{HTML}{003F5C}
\definecolor{ledrcol}{HTML}{58508D}
\definecolor{hecol}{HTML}{BC5090}
\definecolor{hedrcol}{HTML}{FF6361}
\definecolor{ltmcol}{HTML}{FFA600}

\begin{document}

\title{What can CMB observations tell us about the neutrino distribution function?}

\author{James Alvey}
\email{j.b.g.alvey@uva.nl}
\thanks{ORCID: \href{https://orcid.org/0000-0003-2020-0803}{0000-0003-2020-0803}}
\affiliation{GRAPPA Institute, Institute for Theoretical Physics Amsterdam,\\
University of Amsterdam, Science Park 904, 1098 XH Amsterdam, The Netherlands}

\author{Miguel Escudero}
\email{miguel.escudero@tum.de}
\thanks{ORCID: \href{https://orcid.org/0000-0002-4487-8742}{0000-0002-4487-8742}}
\affiliation{Physik-Department, Technische Universit{\"{a}}t, M{\"{u}}nchen, James-Franck-Stra{\ss}e, 85748 Garching, Germany}

\author{Nashwan Sabti}
\email{nashwan.sabti@kcl.ac.uk}
\thanks{ORCID: \href{https://orcid.org/0000-0002-7924-546X}{0000-0002-7924-546X}}
\affiliation{Department of Physics, King's College London, Strand, London WC2R 2LS, UK}

\preprint{TUM-HEP-1375/21, KCL-2021-87}

\begin{abstract}

\noindent Cosmic Microwave Background (CMB) observations have been used extensively to constrain key properties of neutrinos, such as their mass. However, these inferences are typically dependent on assumptions about the cosmological model, and in particular upon the distribution function of neutrinos in the early Universe. In this paper, we aim to assess the full extent to which CMB experiments are sensitive to the shape of the neutrino distribution. We demonstrate that Planck and CMB-S4-like experiments have no prospects for detecting particular features in the distribution function. Consequently, we take a general approach and marginalise completely over the form of the neutrino distribution to derive constraints on the relativistic and non-relativistic neutrino energy densities, characterised by $N_\mathrm{eff} = 3.0 \pm 0.4$ and $\rho_{\nu,0}^{\rm NR} < 14 \, \mathrm{eV}\,\mathrm{cm}^{-3}$ at 95\% CL, respectively. The fact that these are the only neutrino properties that CMB data can constrain has important implications for neutrino mass limits from cosmology. Specifically, in contrast to the $\Lambda$CDM case where CMB and BAO data tightly constrain the sum of neutrinos masses to be $\sum m_\nu < 0.12 \, \mathrm{eV}$, we explicitly show that neutrino masses as large as $\sum m_\nu \sim 3 \, \mathrm{eV}$ are perfectly consistent with this data. Importantly, for this to be the case, the neutrino number density should be suitably small such that the bound on $\rho_{\nu,0}^\mathrm{NR} = \sum m_\nu n_{\nu,0}$ is still satisfied. We conclude by giving an outlook on the opportunities that may arise from other complementary experimental probes, such as galaxy surveys, neutrino mass experiments and facilities designed to directly detect the cosmic neutrino background.

\vspace*{5pt} \noindent \textbf{\texttt{GitHub}}: Parameter files for MCMC analysis and code to reproduce all plots can be found \href{https://github.com/james-alvey-42/DistNuAndPtolemy}{here}.
\end{abstract}

\maketitle
\hypersetup{
  linkcolor  = linkcolor,
  citecolor  = linkcolor,
  urlcolor   = linkcolor,
  colorlinks = true
}

\section{Introduction}\label{sec:Intro}

\noindent Neutrino masses represent the only laboratory evidence for physics beyond the Standard Model of Particle Physics (SM)~\cite{GonzalezGarcia:2002dz,Strumia:2006db}. Although neutrino oscillation experiments have accurately measured the mass-squared differences between the three active neutrinos~\cite{Esteban:2020cvm,deSalas:2020pgw,Capozzi:2021fjo}, at present, we still do not actually know what the absolute mass of any neutrino is. Currently, the best laboratory bound on the neutrino mass comes from the KATRIN experiment that reports $m_{\nu_e} < 0.8\,\text{eV}$ at 90\% CL~\cite{Aker:2019uuj,Aker:2021gma}. In addition, and also from the laboratory, we know from the KamLAND-Zen experiment that the effective Majorana neutrino mass is bounded to be $m_{\beta \beta} < (0.061-0.165)\,\text{eV}$ at 90\% CL~\cite{KamLAND-Zen:2016pfg}. This implies that if neutrinos are Majorana particles, they should likely have a mass $m_{\nu} < 0.48\,\text{eV}$\footnote{In several scenarios beyond the Standard Model, there are other potential contributions to $m_{\beta \beta}$, see e.g.~\cite{DellOro:2016tmg,Dolinski:2019nrj}. Such contributions could interfere destructively with the one arising from the active neutrinos and lead to a situation where active Majorana neutrinos have a mass $m_{\nu} \gtrsim 0.48\,\text{eV}$.}~\cite{KamLAND-Zen:2016pfg}. Here $m_{\nu_e}^2 \equiv \sum_{i}{|U_{ei}|^2 m_i^2}$, where $U_{ei}$ are leptonic mixing matrix elements and $m_i$ are the masses of the neutrino mass eigenstates. In addition, $m_{\beta\beta} = \left|\sum_i{m_iU_{ei}^2}\right|$ is the effective Majorana mass relevant to neutrinoless double beta-decay experiments.

Massive neutrinos have important cosmological implications~\cite{Dolgov:2002wy,Lesgourgues:2006nd,Lesgourgues:2013sjj}, and as such, cosmological observations have been used to set relevant constraints on their properties. In particular, within the standard cosmological model ($\Lambda$CDM), the Planck collaboration reports $\sum m_\nu = \sum_{i}{m_i} < 0.12\,\text{eV}$ at 95\% CL~\cite{Aghanim:2018eyx} (see also~\cite{Vagnozzi:2017ovm,RoyChoudhury:2019hls}) by using very precise observations of the Cosmic Microwave Background (CMB) together with Baryon Acoustic Oscillations (BAO) data~\cite{Beutler:2011hx,Ross:2014qpa,Alam:2016hwk}. This cosmological bound is very important, because \textit{i)} it is based on linear cosmology, namely, it is derived in a regime for which the growth of cosmological perturbations is under full theoretical control, see e.g.~\cite{Dodelson:2003ft}, and \textit{ii)} it is significantly more stringent than current laboratory constraints on $m_\nu$. However, it is important to emphasise that \emph{cosmological constraints on the neutrino mass are cosmological-model dependent}. In particular, in order to derive this Planck constraint, one assumes the standard Big-Bang model expectation that neutrinos decoupled from the rest of the primordial plasma at a temperature of $T \sim 2\,\text{MeV}$, and that they follow a frozen Fermi-Dirac distribution with $T_\nu/T_\gamma \simeq (4/11)^{1/3}\simeq 0.71$~\cite{Kolb:1990vq,Gorbunov:2011zz}. Given the elusive nature of neutrinos and the relevance of cosmological constraints on the neutrino mass, it is pertinent to explore alternatives to this assumption.

In this work, we revisit the idea that neutrinos may not follow a Fermi-Dirac distribution and study the CMB implications of such a scenario. Here, we take a general perspective, and aim to understand the full extent to which CMB observations are sensitive to features in the neutrino distribution function. In fact, as we explicitly demonstrate, current CMB observations are only really sensitive to the energy density of neutrinos, both when they are ultra-relativistic, as parameterised by $N_{\rm eff}$, and when they are non-relativistic, which we denote by $\rho_\nu^{\rm NR}$ (or equivalently $\Omega_\nu h^2$). Importantly, this data-driven result is directly relevant to cosmological constraints on the neutrino mass. In particular, since current CMB data are only sensitive to $\rho_\nu^{\rm NR} = \sum m_\nu n_\nu$, where $n_\nu$ is the neutrino number density, it follows immediately that cosmologies where the number density of neutrinos is smaller than the one predicted in the Standard Model can lead to a (substantial) relaxation of the cosmological neutrino mass bound. This is particularly timely, as the KATRIN experiment is running and should achieve a sensitivity of $m_\nu < 0.2\,\text{eV}$ at 90\% CL within the next four years~\cite{KATRIN:2001ttj}. If a neutrino mass detection is reported by the KATRIN collaboration, neutrinos with a distribution function that deviates from the one in the SM would represent a clear possibility to reconcile precise cosmological observations and laboratory measurements that would appear contradictory within the context of $\Lambda$CDM.

Before continuing, we note that several aspects of this possibility have been studied in the literature~\cite{Cuoco:2005qr,deSalas:2018idd,Oldengott:2019lke,Renk:2020hbs}. In particular, back in 2005, Ref.~\cite{Cuoco:2005qr} considered the cosmological implications of Fermi-Dirac neutrinos supplemented by an additional non-standard neutrino population. More recently, Ref.~\cite{deSalas:2018idd} examined neutrinos with a distribution function that interpolates between Fermi-Dirac and Bose-Einstein, Ref.~\cite{Oldengott:2019lke} explored the implications of scenarios with mild distortions to the Fermi-Dirac spectrum, and Ref.~\cite{Renk:2020hbs} investigated scenarios where neutrinos follow a Fermi-Dirac distribution but with a lower temperature than the one expected in the Standard Model. These references highlighted the role of the neutrino distribution function on the neutrino mass bound, but typically focused on individual examples that are relatively close to the Fermi-Dirac one. In contrast, in this work, we take a broader perspective and consider a wider array of possible neutrino distribution functions, and explicitly demonstrate that the CMB is not sensitive to particular features of the distribution. We will focus on the non-relativistic neutrino energy density as the key quantity in this regard, since it is already well known that constraints on $N_\mathrm{eff}$ are insensitive to the precise form of the neutrino distribution function~\cite{Lesgourgues:2013sjj}.

\begin{figure*}[t!]
    \centering
    \includegraphics[width=0.7\linewidth]{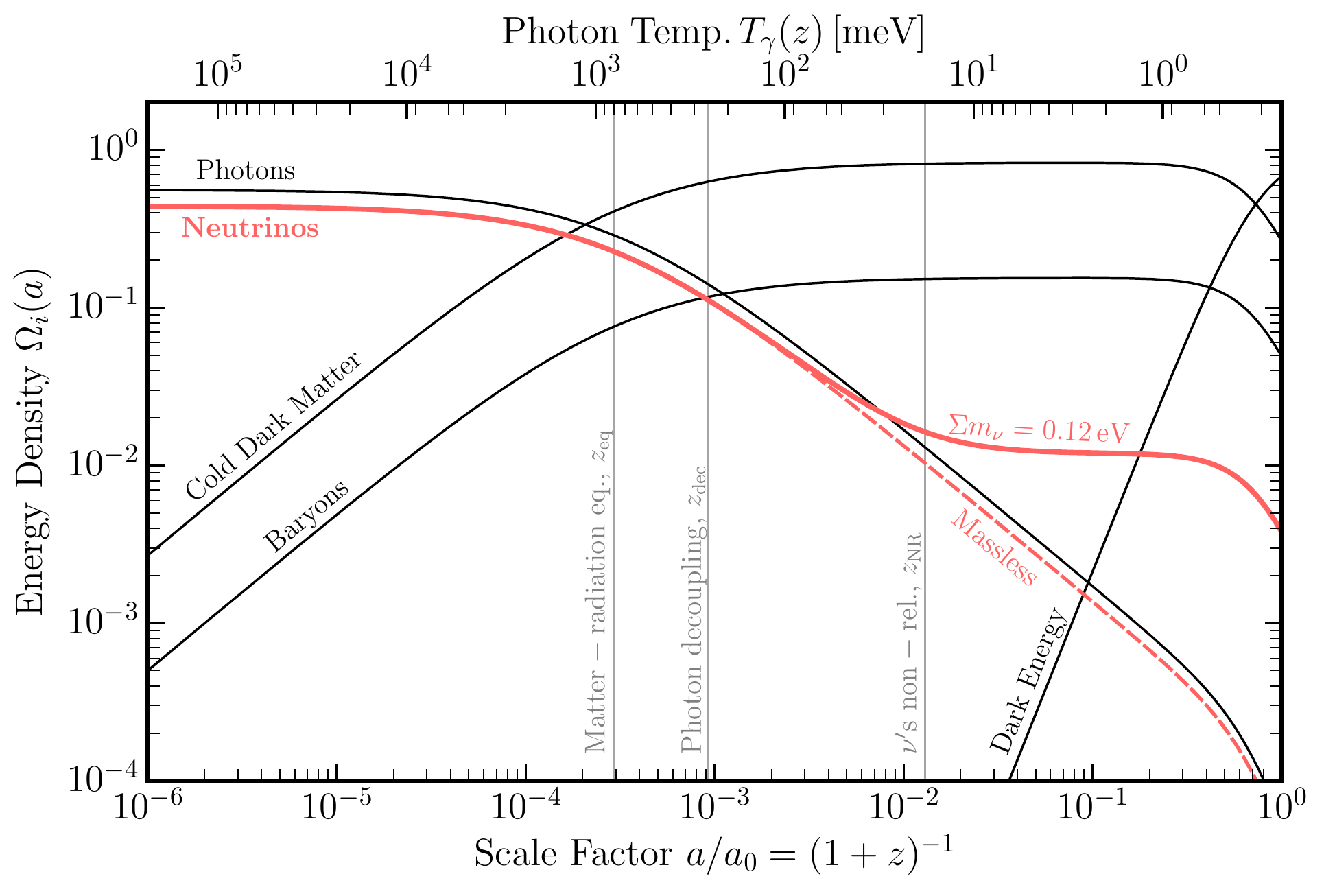}
    \caption{Evolution of the energy density in neutrinos (red), photons, cold dark matter, baryons and dark energy (black) within $\Lambda$CDM. The difference between the massive (solid) and massless (dashed) neutrino curves highlights that only the former makes the transition from a relativistic, radiation-like species to a matter-like one. Note that we have assumed that the neutrino masses are degenerate, $m_{i} = (\sum m_\nu)/3$ in making this figure.}
    \label{fig:energy_densities}
\end{figure*}

Whilst the main focus of our study is the effect of the neutrino distribution function on CMB observations, Big Bang Nucleosynthesis (BBN) can also be used to place constraints on the neutrino distribution function (albeit in a different epoch). More explicitly, the synthesis of the primordial light elements requires the presence of neutrinos for two main reasons (see e.g.~\cite{Sarkar:1995dd}). Firstly, neutrinos are directly involved in the inter-conversion of protons and neutrons in the primordial plasma (via e.g. $\nu_e + n \leftrightarrow p + e^-$), and secondly, the energy density in neutrinos impacts the expansion rate of the Universe in this early epoch. As such, BBN has been used to place constraints on the neutrino distribution, see e.g.~\cite{Cucurull:1995bx,Dolgov:2005mi,Iizuka:2014wma,deSalas:2018idd}, and on the abundance of relativistic (neutrino) species in the early Universe, see~\cite{Pisanti:2020efz,Fields:2019pfx,Pitrou:2018cgg} for the latest analyses. The key takeaway messages from these studies are: \emph{i)} that successful nucleosynthesis requires the presence of neutrinos at the time of proton-to-neutron freeze-out ($t\sim 1\,\text{s}, \, T \sim 0.7\,\text{MeV}$), \emph{ii)} that these neutrinos should roughly interact at the same rate with protons and neutrons as expected in the standard scenario, and \emph{iii)} that the number of ultra-relativistic neutrino species contributing to the expansion of the Universe at the time should be $N_{\rm eff} \simeq 3 \pm 0.3$. In other words, BBN tells us that neutrinos should have been present in the first few minutes of the Universe with broadly the properties expected in the Standard Model. 

However, there is a simple reason why, in spite of the constraints from BBN, it is important to independently consider the implications of non-standard neutrino distributions on the CMB. Specifically, it is the fact that BBN probes the neutrino distribution at $t \lesssim 3\,\text{min}$, leaving plenty of time after which the neutrino distribution could have changed significantly. This could have happened, for example, as a result of the interactions between neutrinos and other species beyond the Standard Model~\cite{Farzan:2015pca}. Therefore, although BBN does offer a window to test the neutrino distribution function, there exists a period of time -- after nucleosynthesis is complete -- where changes in the neutrino distribution may be invisible to observations of both the CMB and the light element abundances. Therefore, in this analysis, we will consider at face value the implications of alternative neutrino distribution functions for CMB observations. 

The remainder of this paper is structured as follows: Firstly, in Sec.~\ref{sec:CMBpheno}, we briefly review the cosmological implications of massive neutrinos. Next, in Sec.~\ref{sec:ExDists}, we consider a set of widely different neutrino distribution functions to explicitly show that current CMB observations are really only sensitive to the energy density of neutrinos when they are non-relativistic, and not directly to their mass or distribution. Within the context of these concrete examples, we also explore the optimal sensitivity of Planck and future CMB-S4 experiments to changes in the neutrino distribution. We find that in both cases there is no sensitivity to features in the neutrino distribution function. In Sec.~\ref{sec:FullAnalysis}, we use a flexible parametrisation of the neutrino distribution, together with an additional dark radiation component, to find marginalised constraints on the energy density of ultra-relativistic and non-relativistic neutrinos. We also summarise our main results, include a discussion regarding the ability to relax the neutrino mass bound in cosmology, and make a direct comparison to previous works. In Sec.~\ref{sec:BBNModelBuilding}, we explore some of the model building implications from BBN and explain how our bounds can be used to determine whether theoretical scenarios to produce non-standard neutrino distribution functions are viable, given our limits. In Sec.~\ref{sec:Summary}, we present our main conclusions. Finally, in Sec.~\ref{sec:Outlook}, we briefly comment on the potential implications of the neutrino distribution function for measurements of the matter power spectrum with future galaxy surveys. Moreover, we motivate the physics case of searches for non-standard neutrino populations at future terrestrial neutrino experiments, such as PTOLEMY, which we discuss in depth in our recent paper~\cite{Alvey:2021xmq}. Technical details and supplementary results as part of our analysis are provided in the Appendices~\ref{app:supplementary_results_sec3}$-$\ref{app:Transition}.

\section{CMB Implications of Massive Neutrinos}\label{sec:CMBpheno}

\noindent The anisotropies observed in the Cosmic Microwave Background provide crucial information about a number of physical processes, including those concerning neutrinos. A detailed description of the impact of (massive) neutrinos on this picture is already well-established, for which we refer the reader to the following reviews~\cite{Dolgov:2002wy,Lesgourgues:2006nd, Hannestad:2010kz, Lesgourgues:2013sjj, Lattanzi:2017ubx}. Here, we will only briefly summarise this story by introducing a simple timeline that highlights why one should expect observations of the CMB to constrain both the relativistic \emph{and} non-relativistic neutrino energy densities. In particular, this timeline splits into three separate epochs based on the neutrino equation of state, see also Fig.~\ref{fig:energy_densities}.

\begin{itemize}
    \item \emph{Relativistic neutrinos}. After neutrino decoupling, at $T\sim 2 \, \mathrm{MeV}$~\cite{Dolgov:2002wy}, neutrinos are highly relativistic and their energy density contributes directly to the effective number of relativistic species:
    \begin{align}
        \label{eq:Neff}
        N_{\rm eff} \equiv \frac{8}{7}\left(\frac{11}{4} \right)^{4/3} \left( \frac{\rho_\text{rad}-\rho_\gamma}{\rho_\gamma}\right)\ ,
    \end{align}
    where $\rho_\mathrm{rad}$ is the total radiation energy density, $\rho_\gamma$ is the photon energy density and $N_\mathrm{eff}^{\rm SM} \simeq 3.044$~\cite{Bennett:2019ewm,Escudero:2020dfa,Akita:2020szl,Froustey:2020mcq,Bennett:2020zkv,Hansen:2020vgm} within the SM. If neutrinos were massless, this would indeed fix the evolution of their energy density for the rest of the cosmic history. On the other hand, for massive neutrinos, they will transition to being non-relativistic at later times, which at the background level alters their contribution to the expansion rate. In this work, we will allow for the possibility that there is also massless dark radiation present in the Universe, so that $N_\mathrm{eff}$ is then a sum of two terms, one coming from the neutrino sector $N_\mathrm{eff}^\nu$ and the other from the dark radiation $N_\mathrm{eff}^\mathrm{DR}$, $N_\mathrm{eff} \equiv N_\mathrm{eff}^\nu + N_\mathrm{eff}^\mathrm{DR}$. The motivation for this choice stems from the fact that very light species that can act as dark radiation are present in many extensions of the Standard Model, see e.g.~\cite{Allahverdi:2020bys} for a review. As is well known~\cite{Lesgourgues:2013sjj}, the CMB is only sensitive to the sum of the two, since the two species only interact gravitationally and therefore lead to the same impact on the CMB spectra.

    \item \emph{Relativistic to non-relativistic transition}. When the average neutrino momentum drops below its mass, neutrinos become a non-relativistic species. Explicitly, this occurs at a redshift $z_\mathrm{NR}$ satisfying $\overline{p}(z_\mathrm{NR}) = \left(\sum m_\nu\right)/3$, which is given by (see Appendix~\ref{app:Transition} for a derivation):
    \begin{align}
        1 + z_\mathrm{NR} = \frac{8}{21}\left(\frac{11}{4}\right)^{4/3}\frac{\rho_{\nu,0}^\mathrm{NR}}{N^\nu_\mathrm{eff}\rho_{\gamma,0}}\ ,
    \end{align}
    where $\rho_{\gamma,0}$ is the photon energy density and $\rho_{\nu,0}^\mathrm{NR}$ is the non-relativistic neutrino energy density, both evaluated today. The key observation here is that the transition from a radiation component to a matter one depends on the ratio between the energy density in non-relativistic neutrinos today and $N_{\rm eff}^\nu$. In the case of a Fermi-Dirac distribution, this reduces to the well-known relation $1 + z_\mathrm{NR} \approx 1890\left(\frac{1}{3}\sum m_\nu / 1\,\mathrm{eV}\right)$, which can found in e.g.~\cite{Lesgourgues:2013sjj}. Of course, the transition between relativistic to non-relativistic matter is not instantaneous and in reality it happens over some redshift range, see also the right panel in Fig.~\ref{fig:nu_app}.
    
    \item \emph{Non-relativistic neutrinos}. Once neutrinos are non-relativistic, they simply contribute as a matter component of the Universe. In contrast to cold dark matter, however, neutrinos typically have large velocities and therefore have a non-zero free streaming length~\cite{Bond:1980ha}. This effect, depending on the amount that they contribute to the total matter density $\Omega_\mathrm{m} = \Omega_\mathrm{b} + \Omega_\mathrm{cdm} + \Omega_\nu$, can have a significant impact on structure formation. In particular, for fixed $\Omega_\mathrm{m}$, an increase in the energy density in neutrinos tends to suppress the formation of structure at small scales compared to that of a pure cold dark matter cosmology, see e.g.~\cite{Blas:2014hya, Lesgourgues:2013sjj}. 
\end{itemize}
 
\noindent This simple picture illustrates how the dynamics of the expanding Universe depends on the neutrino energy density and its equation of state. This does not explain, however, how Planck CMB data actually constrains the relevant parameters. To do this, one should connect this timeline with the physical quantities extracted from measurements of the Cosmic Microwave Background -- e.g. the angular scales associated to the acoustic sound horizon $\theta_\mathrm{s}$ and the Silk damping scale $\theta_\mathrm{d}$, or the comoving angular diameter distance to the CMB $D_\mathrm{A}$. With this, we can then motivate why we are able to constrain, to some degree, both the relativistic energy density -- parameterised by $N_\mathrm{eff}$ -- as well as the non-relativistic energy density $\rho_\nu^{\mathrm{NR}}$. The mechanisms by which each quantity is measured differ significantly and can be understood as follows. 

\begin{itemize}
\item \emph{Relativistic Energy Density.} There are a number of ways to see the impact of the relativistic energy density in neutrinos on the angular scales ($\theta_\mathrm{s}$, $\theta_\mathrm{d}$) mentioned above. The most common approach involves fixing $\theta_\mathrm{s}$~\cite{Hou:2011ec}\footnote{Keeping all other cosmological parameters constant, except for $N_\mathrm{eff}$, which then fixes the rest of the acoustic peak positions.}, since it is very precisely determined from CMB data. The dominant physical effect of the relativistic energy density in neutrinos is that they contribute to the Hubble rate before recombination. The sensitivity of the CMB to $N_\mathrm{eff}$ then arises from the different dependence of the acoustic sound horizon $r_\mathrm{s}$ and damping scale $r_\mathrm{d}$ on the Hubble rate~\cite{Hou:2011ec}. In other words, since we can measure the ratio $\theta_\mathrm{s}/\theta_\mathrm{d} = (r_\mathrm{s}/D_\mathrm{A})/(r_\mathrm{d}/D_\mathrm{A}) \sim H^{-1/2}$ from the CMB, we are able to obtain information about the early-time Hubble rate, and therefore $N_\mathrm{eff}$. Said differently, if we fix $\theta_\mathrm{s}$, then any change from varying $N_\mathrm{eff}$ will be restricted to the damping tail through $\theta_\mathrm{d}$.\footnote{Note that the CMB damping scale is also sensitive to the electron number density $n_e$, which in turn depends on the primordial helium abundance $Y_\mathrm{P}$ through $n_e \propto (1-Y_\mathrm{P})$. This results in a degeneracy between $N_\mathrm{eff}$ and $Y_\mathrm{P}$, which we do not consider in this work, as we fix $Y_\mathrm{P} = 0.245$~\cite{pdg} throughout our analysis. Nevertheless, given current measurements of $Y_\mathrm{P}$~\cite{Izotov:2014fga,Aver:2015iza,Peimbert:2016bdg,Fernandez:2018,Valerdi:2019beb}, the impact of variations in $Y_\mathrm{P}$ is negligible for Planck CMB observations (see e.g. Fig. 41 in~\cite{Aghanim:2018eyx}).}

\item \emph{Non-relativistic Energy Density.} In contrast to the relativistic case, the leading-order effect of the non-relativistic energy density in neutrinos is via the comoving angular diameter distance $D_\mathrm{A}$. This shifts all angular scales measured in the CMB to lower multipoles~\cite{Lesgourgues:2006nd}. One might naively think that this means that the effect of the non-relativistic neutrino energy density is highly degenerate with other cosmological parameters, such as $H_0$ or $\Omega_\Lambda$. As we discussed in the timeline, however, a very important property of neutrinos is that they have a non-negligible free-streaming length, even after they become non-relativistic. Depending on the value of the ratio $\rho_\nu^\mathrm{NR}/\rho_\mathrm{m} \equiv f_\nu$, this may have a significant effect on the matter power spectrum at small scales. Importantly, the propagation of the CMB photons from the time of last scattering until today is affected by the presence, or absence, of structure along their path~\cite{pdg}. This results in a number of relatively small but important late-time effects on the CMB power spectrum, e.g.  weak lensing~\cite{Lewis:2006fu, Aghanim:2018oex} and the late Integrated Sachs-Wolfe effect~\cite{Sachs:1967,Giannantonio:2008zi}. The fact that each of these effects has a specific scale dependence allows one to (partially) break the degeneracy between cold dark matter, baryons, non-relativistic neutrinos and dark energy that enters in the expression for $D_\mathrm{A}$. For example, the suppression of the matter power spectrum due to massive neutrinos acts to effectively reduce the weak lensing experienced by CMB photons only on angular scales $\theta \lesssim 1^\circ$~\cite{pdg}.
\end{itemize}

\noindent In summary, measurements of the CMB on a wide range of angular scales can probe both the early- and late-time properties of neutrinos, with the key quantities that can be constrained by CMB data being the relativistic and non-relativistic neutrino energy densities. As a consequence, this raises the question of whether we can directly access any information about the form of the neutrino distribution function after the BBN epoch, aside from these integrated properties. In the next section, we will start to address this by considering a set of example distributions which differ greatly in their average momentum and variance. Despite these differences, they are constructed to have the same relativistic and non-relativistic energy densities. We will explicitly show that Planck CMB data cannot tell the difference between these neutrino distributions, and that there are similar prospects for future CMB missions such as CMB-S4. Ultimately, the aim of this exercise is to set up the expectations for the full analysis in Sec.~\ref{sec:FullAnalysis}, where we place constraints on $N_\mathrm{eff}$ and $\rho^\mathrm{NR}_{\nu,0}$, marginalising completely over the neutrino distribution.

\section{The Impact of the Neutrino Distribution Function on the CMB} \label{sec:ExDists}

\noindent Having described the key cosmological implications of massive neutrinos, in this section we will explicitly showcase how inferences of cosmological parameters with current Planck CMB data are insensitive to the specific shape of the neutrino distribution function. To do this, we consider five scenarios in which neutrinos have a drastically different distribution\footnote{Note that these distributions are not meant to be necessarily physically motivated, but instead chosen to precisely highlight how widely different distribution functions can lead to the same results at the level of current and future data.}, i.e., their characteristic average momentum and variance differ greatly, but where the total $N_\mathrm{eff} = N_\mathrm{eff}^\nu + N_\mathrm{eff}^\mathrm{DR}$ and non-relativistic energy density $\rho_{\nu, 0}^\mathrm{NR} = \sum m_\nu n_{\nu, 0}$ are the same. Given this setup and the discussion in the previous section, we expect that any effect on the CMB should be small, despite these differences. We illustrate this in Fig.~\ref{fig:distributions}, along with the evolution of the total energy density in neutrinos and dark radiation, and the implications for the CMB TT power spectrum. More quantitatively, by carrying out a full MCMC analysis using Planck legacy data, we will show that in each case the neutrino mass limit is altered in such a way to give the same bound on the non-relativistic neutrino energy density. This will then highlight that Planck CMB data cannot distinguish between widely different distribution functions, provided that the non-relativistic energy density $\rho_{\nu,0}^{\rm NR}$ and total $N_\mathrm{eff}$ are the same. With this in mind, we will finally show that there are also only limited prospects for future experiments, such as CMB-S4, to detect differences in the neutrino distribution function.

\subsection{Choice of Distributions}

\noindent In practice, we use two specific parameterisations of the neutrino distribution:
\begin{align}
    \label{eq:dist_thermal}
    \mathbf{Fermi\!-\!Dirac} &= 
    \begin{cases}
    f_\nu(q_\nu) = \left(e^{q_\nu} + 1\right)^{-1}\,,
    \end{cases}
    \\
    \label{eq:dist_Gaussian}
    \mathbf{Gaussian} &= 
    \begin{cases}
    f_\nu(q_\nu | N^\nu_\mathrm{eff}, y_*, \sigma_*) = \\A(N_\mathrm{eff}^\nu, y_*, \sigma_*)\exp{\left(-\frac{(q_\nu - y_*)^2}{2\sigma_*^2}\right)}\,,
    \end{cases}
\end{align}
where $q_\nu = p_\nu/T_\nu$ is the comoving momentum. We note that in the Gaussian case, $T_\nu$ is not a well-defined quantity, and thus we fix it to $T_\gamma/T_\nu = 1.39578$~\cite{Escudero:2020dfa} without any loss of generality. The variables $y_*$ and $\sigma_*$ are the free parameters of the Gaussian distribution and have the interpretation of the average momentum and momentum variance, respectively (see Sec.~\ref{sec:FullAnalysis} for further details on this aspect). In addition, the amplitude $A$ in Eq.~\eqref{eq:dist_Gaussian} is tuned to give the correct input value of $N_\mathrm{eff}^\nu$, which can be computed using:
\begin{equation}\label{eq:Neff_distribution}
    N_{\mathrm{eff}}^\nu = \frac{360}{7 \pi^4} \left(\frac{11}{4}\right)^{4/3} \left(\frac{T_\nu}{T_\gamma}\right)^4 \int_0^\infty{\mathrm{d}q_\nu \, q_\nu^3 f_\nu(q_\nu)}\ ,
\end{equation}
where within the Standard Model this evaluates to $N_\mathrm{eff}^\nu = 3 \left(11/4\right)^{4/3} \left(T_\nu/T_\gamma\right)^4 = 3.044$. Similarly, the number density can be obtained from:
\begin{equation}\label{eq:num_nu}
    \frac{n_\nu}{n_\nu^{\mathrm{FD}}} = \frac{2}{3 \zeta(3)} \int_0^{\infty}{\mathrm{d}q_\nu \, q_\nu^2 f_\nu(q_\nu)}\ ,
\end{equation}
where $n_{\nu,0}^\mathrm{FD} = 3\zeta(3)T_{\nu,0}^3/(2\pi^2) \simeq 114\,\mathrm{cm^{-3}}$ is the neutrino number density for a single flavour, assuming a Fermi-Dirac distribution. The five different scenarios we consider are then as follows:
\begin{itemize}
    \item {\color{lcdmcol} \boldmath{$\Lambda$}\textbf{CDM}} The SM case, where neutrinos have a Fermi-Dirac distribution with temperature $T_\nu = T_\gamma/1.39578$. This gives $N_\mathrm{eff} = N_\mathrm{eff}^\nu = 3.044$.
    
    \item {\color{ledrcol} \textbf{L}\boldmath{$\nu$}\textbf{-DR}} A low-energy neutrino population with a Gaussian distribution (where $N_\mathrm{eff}^\nu = 0.5$, $y_* = 0.1$ and $\sigma_* = 0.294218$), complemented with massless dark radiation (DR) to give a total $N_\mathrm{eff} = N_\mathrm{eff}^\nu + N_\mathrm{eff}^\mathrm{DR} = 3.044$.
    
    \item {\color{hecol} \textbf{H}\boldmath{$\nu$}} A high-energy neutrino population with a Gaussian distribution, where $N_\mathrm{eff}^\nu = 3.044$, $y_* = 30$ and $\sigma_* = 4.82113$.
    
    \item {\color{hedrcol}  \textbf{H}\boldmath{$\nu$}\textbf{-DR}} A high-energy neutrino population with a Gaussian distribution (where $N_\mathrm{eff}^\nu = 1.5$, $y_* = 3$ and $\sigma_* = 8.82654$), complemented with massless dark radiation.
    
    \item {\color{ltmcol} \textbf{LT+Mid}} A mid-energy neutrino population with a Gaussian distribution (where $N_\mathrm{eff}^\nu =  2.3139$, $y_* = 3.5$ and $\sigma_* = 0.508274$), together with a low-temperature population that has a Fermi-Dirac distribution with $T_\nu = T_\gamma/2$.
\end{itemize}

\begin{figure*}[t]
    \centering
    \includegraphics[width=0.92\linewidth]{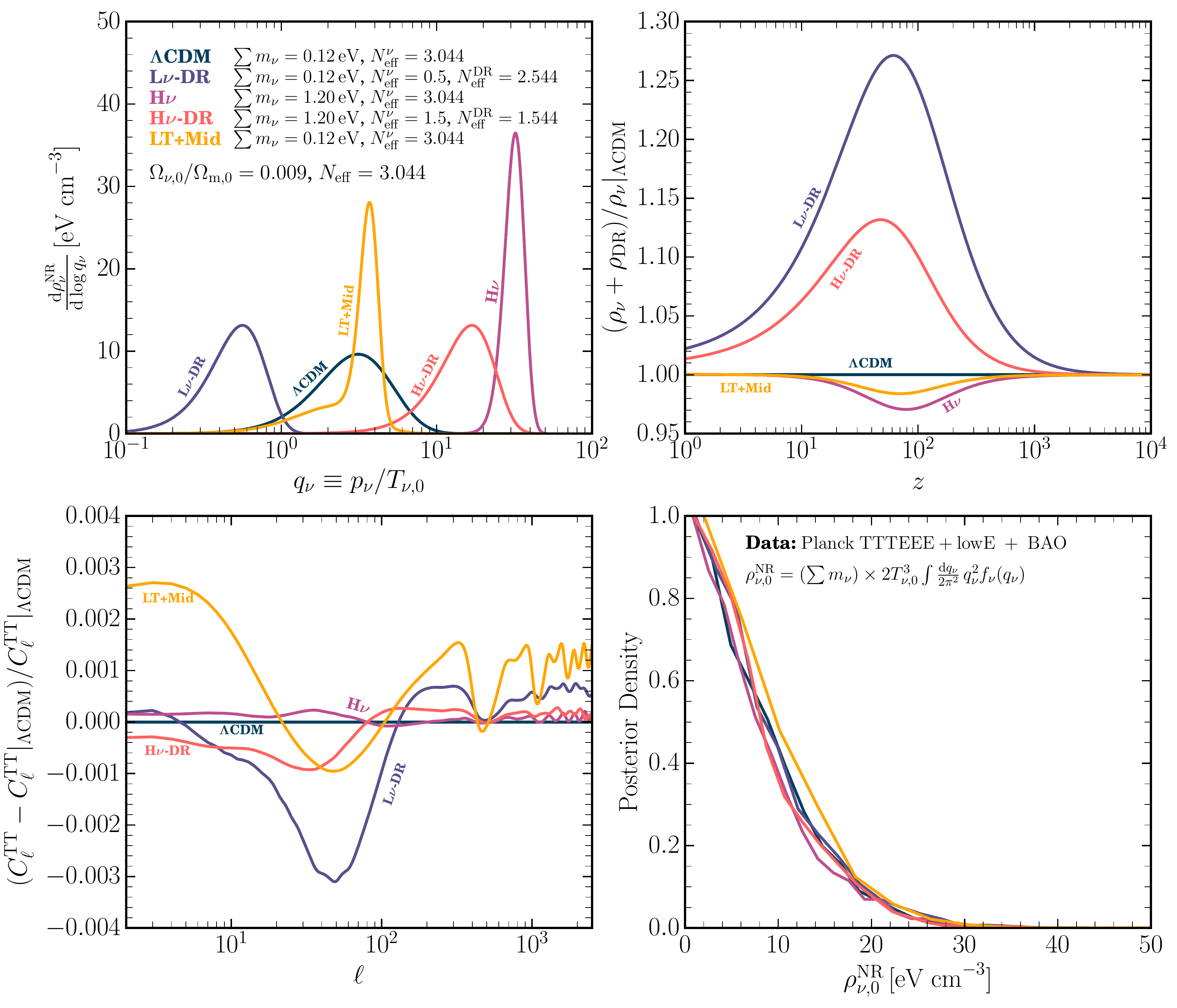}
    \caption{\textit{Upper-left panel:} Differential non-relativistic energy densities as a function of comoving momenta for a common $\rho_{\nu,0}^{\rm NR}$ and $N_{\rm eff} = 3.044$. Note that the area under each curve in this panel is the same. \textit{Upper-right panel:} Energy density evolution of the neutrinos and dark radiation as a function of redshift. \textit{Lower-left panel:} Relative difference in the TT CMB power spectrum between the various cases. For comparison, note that the Planck statistical error bars at $\ell \sim 1000$ are $\Delta C_\ell / C_\ell \sim 0.8\%$ (for a $\Delta \ell = 30$ power band), so the magnitude of the effect shown here is significantly smaller for all angular scales resolved by Planck.
    \textit{Lower-right panel:} Marginalised constraints on $\rho_{\nu,0}^{\rm NR}$ from a full Planck legacy analysis for each of the cases. We can clearly appreciate that Planck data cannot distinguish these rather different scenarios.}
    \label{fig:distributions}
\end{figure*}

\noindent The differential non-relativistic energy density for all five of these non-standard neutrino distributions are shown in the upper-left panel of Fig.~\ref{fig:distributions}. From this panel, it can clearly be seen that, while all five distributions give the same non-relativistic energy density $\rho_{\nu,0}^\mathrm{NR}$ (found by integrating over $\log q_\nu$ in this figure), the neutrino populations are significantly different from the thermal Fermi-Dirac distribution within the SM. This is important given the context of previous works on non-standard neutrino distributions in cosmology, where the departures from a thermal Fermi-Dirac distributions were rather mild.

In addition to the distributions, the top-right panel of Fig.~\ref{fig:distributions} shows the evolution of the neutrino and dark-radiation energy densities with respect to the one within the SM. We see how both the {\color{ledrcol} \textbf{L}\boldmath{$\nu$}\textbf{-DR}} and {\color{hedrcol}  \textbf{H}\boldmath{$\nu$}\textbf{-DR}} cases cause an increase of the energy-density ratio, which is simply because the dark-radiation energy density is still relatively large, even at the time when neutrinos become non-relativistic. On the other hand, in the {\color{ltmcol} \textbf{LT+Mid}} and {\color{hecol} \textbf{H}\boldmath{$\nu$}} cases, the ratio decreases, which is due to the fact that the neutrinos here have a higher average momentum than in the SM case and become non-relativistic later. 

Of course, since the precise energy density evolution affects the Hubble rate, this change in the evolution of the neutrino energy density does have an impact on the CMB temperature power spectra, which we plot in the lower-left panel of Fig.~\ref{fig:distributions}. On the other hand, given that Planck errors bars range between $3\%-0.7\%$ for low $\ell$ and high $\ell$, respectively, we can immediately appreciate the fact that the changes caused by the different distributions are highly unlikely to have observational consequences, at least at the level of current data. Plots for the neutrino equation of state, unlensed TT power spectrum and evolution of the non-relativistic neutrino energy density are provided in  App.~\ref{app:supplementary_results_sec3}.

\subsection{CMB Analysis for Fixed Distributions}
\noindent Of course, we would like to explicitly test the hypothesis that there is no real distinguishable feature in the neutrino distribution function given current state-of-the-art Planck CMB data. To do so, we implement each of these distributions in the cosmological Boltzmann solver \texttt{CLASS}~\cite{Lesgourgues:2011re, Blas:2011rf} and perform a baseline Planck 2018 TTTEEE+lowE + BAO analysis (the full details and datasets are described in Sec.~\ref{sec:FullAnalysis}) using the MCMC sampler \texttt{MontePython}~\cite{Audren:2012wb, Brinckmann:2018cvx}. To understand the outcome of this computation, one should bear in mind that in reality, the parameter that is varied is the neutrino mass, however, the non-relativistic energy density can be directly computed using $\rho_\nu^\mathrm{NR} = \sum m_\nu n_\nu$, where $n_\nu$ is fixed (but different) for each distribution. We use a very wide prior on the individual neutrino mass of $m_\nu \in [0, 10]\, \mathrm{eV}$. Of course, we know that from the laboratory $m_\nu < 0.8\,{\rm eV}$, but since the number density of neutrinos varies up to a factor of 10 between the distribution functions, we need to use a large prior in order to find meaningful cosmological bounds on $\rho^{\rm NR}_{\nu,0}$. 

\begin{figure*}[t]
    \vspace*{-0.3cm}
    \centering
    \includegraphics[width=0.86\linewidth]{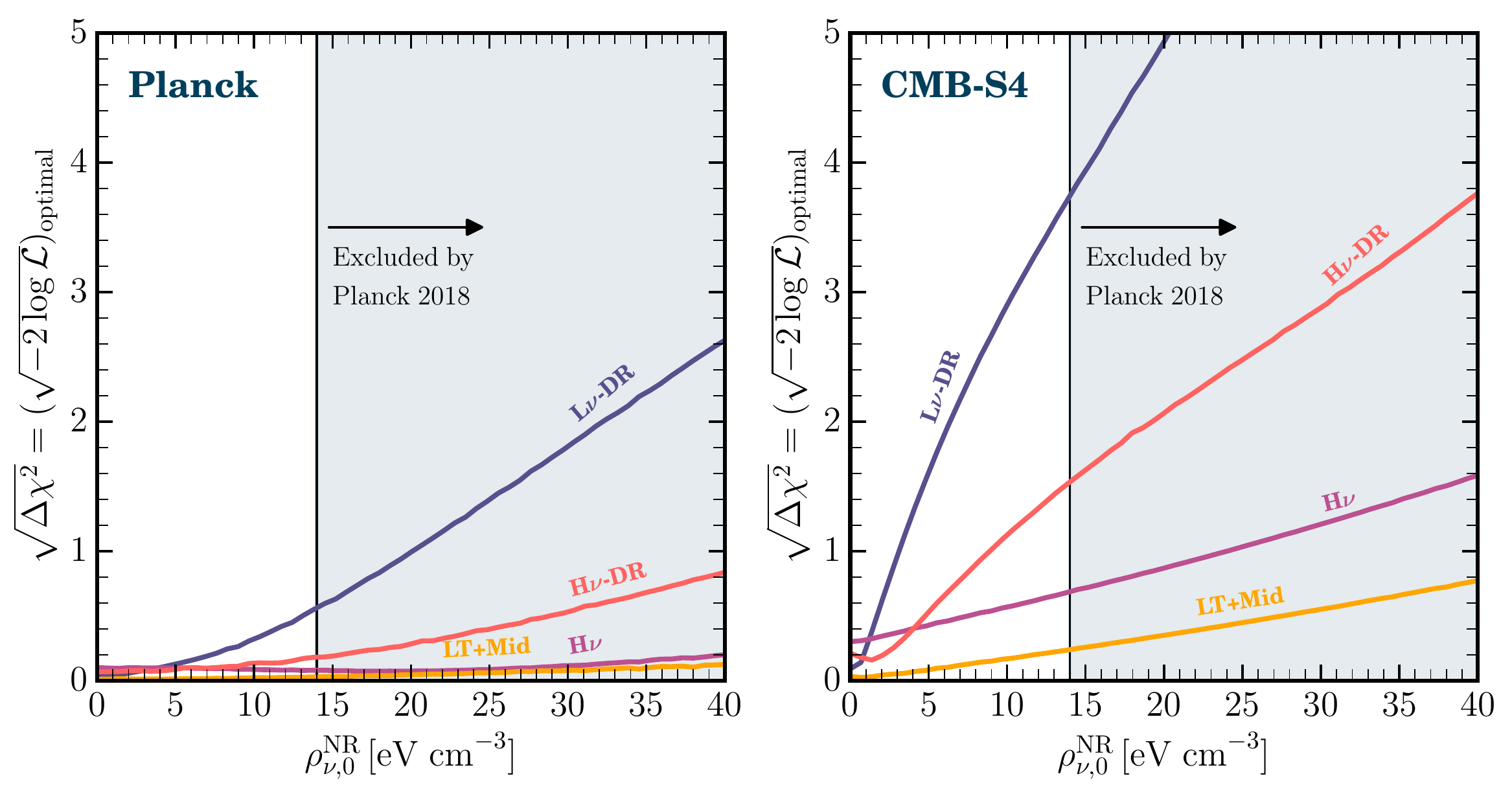}
    \caption{Estimate of the optimal sensitivity of CMB experiments (\emph{Left:} Planck, \emph{Right:} CMB-S4) to the precise form of the neutrino distribution as a function of the non-relativistic energy density in neutrinos today $\rho_{\nu,0}^{\mathrm{NR}}$. We see that with Planck data there are no prospects for detecting differences in the distribution for energy densities below the limit imposed by a full Planck 2018 analysis. Similarly, whilst the prospects appear somewhat better for CMB-S4, we emphasise that this is the optimal case. Indeed, we have explicitly checked that a full parameter scan would reduce all sensitivities to at least below the $2\sigma$ level for energy densities below the Planck bound. In addition, CMB-S4 experiments should improve the upper bound on $\rho_{\nu,0}^{\mathrm{NR}}$, which would further diminish the ability to distinguish between non-standard neutrino distributions.}
    \label{fig:cmbs4}
\end{figure*}

The 1D posteriors on the non-relativistic neutrino energy density are shown in the lower-right panel of Fig.~\ref{fig:distributions}. We see that in all five cases we obtain the same limits on $\rho_{\nu,0}^\mathrm{NR}$, no matter how different the neutrino distributions are, confirming our previous expectations regarding the insensitivity of current CMB data to deviations in the neutrino distribution function. This allows us to come to our first main conclusion in this paper: \emph{At the level of current Planck CMB data, the bounds on the non-relativistic (and relativistic) energy density in neutrinos are independent of the neutrino distribution function.}

\subsection{Future CMB Data}

\noindent The previous analysis highlights the fact that current CMB data does not have the constraining power to search for particular features in the neutrino distribution function. This raises the question, however, of whether future experiments, such as CMB-S4, could have sensitivity where Planck does not. To try and answer this question, we have performed the following exercise:

\begin{enumerate}
    \item[] \textbf{Step 1.} Imagine that the actual CMB data observed by an experiment (with either Planck- or CMB-S4-like error bars) was that of a $\Lambda$CDM neutrino population with a specific value of the non-relativistic energy density $\rho_{\nu,0}^\mathrm{NR}$. The power spectrum of this type of scenario will act as a fiducial data set in our exercise.
    \item[] \textbf{Step 2.} Now, for each of the other four distributions described above, choose $\sum m_\nu$ so that the energy density (and the total $N_\mathrm{eff}$) match the $\Lambda$CDM case, and compute the CMB temperature power spectrum.
    \item[] \textbf{Step 3.} This computed power spectrum will likely differ from that in \textbf{Step 1}, however given our experience from above, it is not likely to deviate much. As such, one way to estimate the ``optimal" sensitivity of a Planck- or CMB-S4-like experiment is to compute the likelihood $\mathcal{L}$ associated to this prediction given the fiducial data generated in \textbf{Step 1}. If the quantity $-2\ln(\mathcal{L})$ is very small, then it suggests that the given experiment cannot distinguish between the different distribution functions for that value of the non-relativistic energy density.
\end{enumerate}

\noindent Before describing the analysis and the results, we should clarify that this approach represents an ``optimal" sensitivity analysis in a very specific sense. In particular, a rigorous analysis should marginalise this likelihood over \emph{all} other possible values of the cosmological parameters. The result of doing this full computation will very likely reduce the expected sensitivity, possibly by a significant amount. On the other hand, if the sensitivity predicted by the method described above is already very low, then this ``optimal" sensitivity can be regarded as a robust upper bound, and any marginalisation procedure can only decrease it further.

With this caveat in place, we can now turn to the analysis and the results. To implement the scheme above, we use the \texttt{Planck} bluebook likelihood and the \texttt{CMB-S4} mock likelihood  provided by default with \texttt{MontePython}~\cite{Audren:2012wb, Brinckmann:2018cvx} and evaluate them across a grid of non-relativistic energy densities $\rho_{\nu, 0}^\mathrm{NR}$. The results of this are shown in the left (Planck) and right (CMB-S4) panels of Fig.~\ref{fig:cmbs4}, along with a band indicating the current limit imposed by Planck (which we will obtain in the next section). From this, we can clearly see that in the region allowed by current data -- in other words, to the left of the shaded region -- Planck has little to no sensitivity to alternative forms of the distribution function, reconfirming the analysis in the previous section. More interestingly, however, we see that for at least three of the example distributions ({\color{hecol} \textbf{H}\boldmath{$\nu$}}, {\color{ltmcol}\textbf{LT+Mid}} and {\color{hedrcol}\textbf{H}\boldmath{$\nu$}\textbf{-DR}}), CMB-S4 also has a sensitivity below $2\sigma$. The only stand-out case is that denoted {\color{ledrcol} \textbf{L}\boldmath{$\nu$}\textbf{-DR}}, which deserves particular attention, since it appears that CMB-S4 potentially has sensitivity in this regime. To understand whether this is indeed the case, we carried out a full MCMC analysis with an energy density $\rho_{\nu, 0}^\mathrm{NR}$ that saturated the Planck limit -- i.e. we performed the marginalisation procedure mentioned above. We found that the actual sensitivity decreased to below $2\sigma$, bringing it in line with the other cases, and realising our cautionary words above. With this check completed, however, it allows us to come to the second key conclusion regarding the CMB and non-standard distribution functions: 
\emph{Current and future CMB experiments will not be able to distinguish between different neutrino distribution functions, provided that they have the same $N_{\rm eff}$ and $\rho_{\nu,0}^{\rm NR}$.}

\section{Full Analysis and Results}
\label{sec:FullAnalysis}

\noindent We have seen in the last section that five widely different distribution functions cannot be distinguished by Planck data provided that they possess the same $N_{\rm eff}$ and $\rho_{\nu,0}^{\rm NR}$. In this section, we will carry out a full statistical analysis that samples among a large variety of neutrino distribution functions and where all relevant cosmological parameters are varied independently. The aim of this is to derive robust limits on the non-relativistic and relativistic neutrino energy densities that are marginalised over the neutrino distribution function. As in Sec.~\ref{sec:ExDists}, we will also allow for the possibility that there is additional dark radiation, parameterised by its contribution to the energy density in relativistic species $N_\mathrm{eff}^\mathrm{DR}$. After presenting the data analysis strategy and the full set of results, we will compare our approach to the ones in previous literature that explored non-standard neutrino distribution functions.

\begin{figure*}
    \centering
    \includegraphics[width=0.71\linewidth]{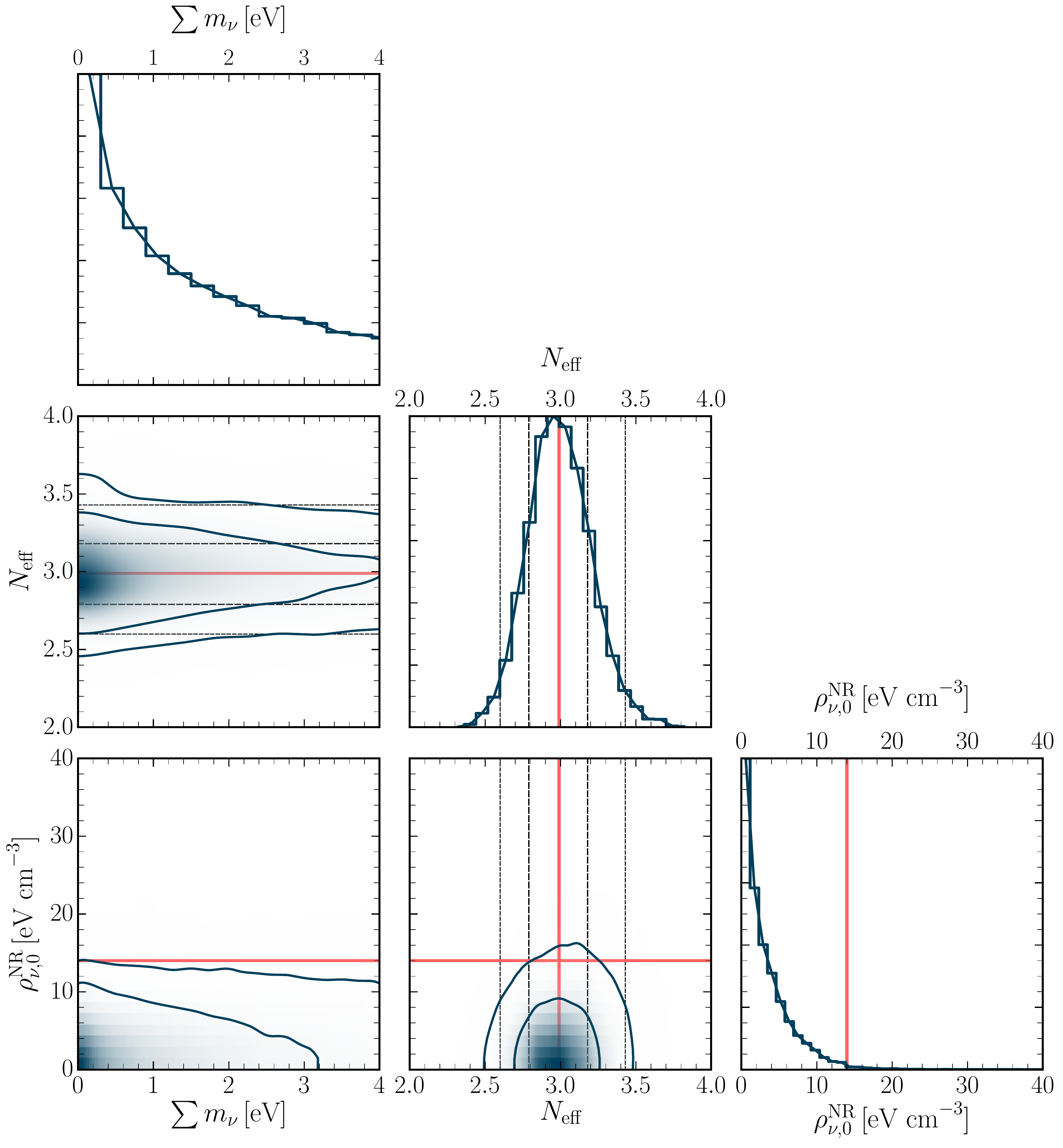}
    \caption{The 1D and 2D confidence intervals for the sum of neutrino masses, relativistic degrees of freedom, and non-relativistic neutrino energy density from our analysis. Note in particular that $\sum m_\nu$ is effectively unconstrained, exploring the full prior range, whilst $\rho_{\nu, 0}^\mathrm{NR} < 14 \, \mathrm{eV}\,\mathrm{cm}^{-3}$ at 95\% CL. Similarly, as expected, $N_\mathrm{eff} = 3.0 \pm 0.4$, also at 95\% CL, in line with the analysis in $\Lambda$CDM.}
    \label{fig:triangle}
\end{figure*}

\subsection{Choice of Distribution Function} 

\noindent We first need to choose a suitably general form for the neutrino distribution function that allows us to independently vary $N_\mathrm{eff}^\nu$, $\rho_{\nu, 0}^{\mathrm{NR}}$ (or equivalently the number density $n_{\nu, 0}$) and the sum of the neutrino masses $\sum{m_\nu}$, in addition to $N_\mathrm{eff}^\mathrm{DR}$. Throughout this study, we consider degenerate neutrinos as we are mostly interested in regions of parameter space where $\sum m_\nu$ is much larger than the neutrino mass splittings. As we argued above, we expect the constraints on $N_{\rm eff}$ and $\rho_{\nu,0}^{\rm NR}$ to be insensitive to the particular form of $f_\nu(q_\nu)$. In order to carry out a concrete analysis, however, a choice must be made and we make use of the Gaussian distribution shown in the previous section, where $f_\nu(q_\nu)$, $N_{\rm eff}$ and $n_{\nu,0}$ are given in Eqs.~\eqref{eq:dist_Gaussian}$-$\eqref{eq:num_nu}, respectively.

While other choices of distribution functions may capture certain physical scenarios more clearly, see e.g.~\cite{Cuoco:2005qr,deSalas:2018idd,Oldengott:2019lke,Renk:2020hbs}, there are two key benefits for this analysis that come with our choice of a Gaussian distribution. Firstly, it allows us to obtain full analytic expressions for $N_\mathrm{eff}^\nu$ and the non-relativistic energy density $\rho_\nu^{\mathrm{NR}} = \sum m_\nu n_\nu$. Secondly, the parameters $y_*$ and $\sigma_*$ can be easily interpreted in terms of their impact on the average momentum and number density of neutrinos. In this regard, for fixed $\sigma_*$ and $N_\mathrm{eff}^\nu$, an increase in $y_*$ directly corresponds to an increase in the average momentum $\overline{p}$ of neutrinos and subsequently a decrease in the number density, since $N_\mathrm{eff}^\nu \propto \overline{p}\,n_\nu$. Likewise, $\sigma_*$ makes the distribution broader and thus includes neutrinos with higher momenta. From Eq.~\eqref{eq:Neff_distribution}, it is clear that the higher momentum neutrinos contribute more to $N_\mathrm{eff}^\nu$ than the lower momentum ones. So again, if we fix $N_\mathrm{eff}^\nu$ and $y_*$, then an increase in $\sigma_*$ will also lead to a decrease in the number density.

\subsection{Cosmological Data Analysis}
\noindent To accurately account for the impact of massive neutrinos on the CMB, we use a slightly modified version of the publicly available Boltzmann code \texttt{CLASS}~\cite{Lesgourgues:2011re, Blas:2011rf}. In particular, we model the neutrinos as a non-cold dark matter component with a phase-space distribution of the form in Eq.~\eqref{eq:dist_Gaussian}, providing $N_\mathrm{eff}^\nu$, $\sum m_\nu$, $y_*$ and $\sigma_*$ as input parameters. We also, in addition, consider massless dark radiation contributing to $N_\mathrm{eff}^\mathrm{DR}$ in the sense described above.  

\newpage

We make use of the Planck legacy data, and in particular the TTTEEE+lowE likelihood~\cite{Aghanim:2019ame}, along with BAO data from the 6dF survey~\cite{Beutler:2011hx}, the MGS sample of the SDSS survey~\cite{Ross:2014qpa}, and the DR12 BAO data from the BOSS survey~\cite{Alam:2016hwk}. We then utilise the Markov-Chain Monte-Carlo code \texttt{Monte Python}~\cite{Audren:2012wb, Brinckmann:2018cvx} to vary all relevant cosmological and nuisance parameters as done in the fiducial Planck analysis. This ensures that we explore all relevant degeneracies between the parameters at the level of the CMB fit. Finally, we take the standard priors from the baseline Planck 2018 analysis~\cite{Aghanim:2019ame}, as well as conservative priors on $N_\mathrm{eff}^\nu \in [0, 10]$, $N_\mathrm{eff}^\mathrm{DR} \in [0, 10]$, $\sum m_\nu \in [0, 6]\,\mathrm{eV}$, $\log_{10}y_* \in [-1, 1.5]$ and $\log_{10}\sigma_* \in [-1, 1.5]$. The logarithmic priors on $y_*$ and $\sigma_*$ ensure that we uniformly sample the number density ratio $n_\nu / n_\nu^{\mathrm{FD}}$, as opposed to biasing the sampling towards values less than unity.

\subsection{Results}\label{sec:Results}

\noindent The main results of our analysis are shown in Fig.~\ref{fig:triangle}, where we confirm the expectations of Sec.~\ref{sec:ExDists}. In particular, we find the following constraints on the relativistic and non-relativistic energy densities at 95\% confidence level:
\begin{align}
    \label{eq:bounds}
    &N_\mathrm{eff} = N_\mathrm{eff}^\nu + N_\mathrm{eff}^\mathrm{DR} = 3.0 \pm 0.4, \nonumber \\
    &\rho_{\nu, 0}^{\mathrm{NR}} < 14 \, \mathrm{eV} \, \mathrm{cm}^{-3}\ .
\end{align} 
This latter bound can be recast in terms of other commonly-used density parameters and fractions: $\Omega_{\nu,0} < 0.0029$, $\Omega_{\nu,0}h^2 < 0.0013$ or $f_{\nu,0} = \Omega_{\nu,0}/\Omega_{\rm m,0} < 0.009$. It is interesting to note that the constraint on the non-relativistic energy density is only an upper bound. This is another way to illustrate the fact that Planck CMB data is compatible with neutrinos being massless~\cite{Aghanim:2018eyx}. Also shown in Fig.~\ref{fig:triangle} is the posterior distribution for the sum of neutrino masses $\sum m_\nu$. We see that within our setup, where we allow the number density of neutrinos to vary, the bound on the neutrino masses is naturally relaxed as compared to the case where they have a Fermi-Dirac distribution with the expected temperature in the Standard Model. Indeed, we see that in our case the upper limit on the sum of the neutrino masses extends all the way to the end of the prior range and is therefore essentially unbounded. For comparison, in the case of the Fermi-Dirac distribution with $T_\nu^{\rm SM}$, this limit instead reads $(\sum m_\nu)^\mathrm{FD} < 0.12 \, \mathrm{eV}$~\cite{Aghanim:2018eyx}. Importantly, however, we can see that this limit is nonetheless fully compatible with our bound on the non-relativistic energy density by considering $(\sum m_\nu)^{\mathrm{FD}} < \rho_{\nu,0}^\mathrm{NR}/n_{\nu,0}^{\mathrm{FD}} = 14 \, \mathrm{eV}\,\mathrm{cm}^{-3} / 114 \, \mathrm{cm}^{-3} \sim 0.12 \, \mathrm{eV}$. A larger triangle plot with more parameters can be found in App.~\ref{app:fullsetofposteriors}.

So far, all of our bounds have been based on cosmological data alone, including the one on $\rho_{\nu, 0}^\mathrm{NR} = \sum m_\nu n_{\nu, 0}$. On the other hand, oscillation experiments show that there is a lower limit on the sum of neutrino masses of $\sum m_\nu \geq 0.058\, \mathrm{eV}$~\cite{Esteban:2020cvm,deSalas:2020pgw,Capozzi:2021fjo}. As such, by fixing the sum of neutrino masses to this minimum value, the bound on $\rho_{\nu, 0}^\mathrm{NR}$ automatically translates into an upper limit on the neutrino number density of $n_{\nu, 0} \lesssim 241 \, \mathrm{cm}^{-3}$ for a single neutrino species. Importantly, this constraint can be considered as a robust upper bound on the number density, irrespective of the value of the neutrino mass, or the form of the distribution function.

\subsection{Comparison with Previous Studies}

\noindent The impact of non-standard neutrino distributions on the CMB has been studied before in~\cite{Cuoco:2005qr,deSalas:2018idd,Oldengott:2019lke,Renk:2020hbs}. Here, we will briefly discuss the approaches employed in each of these references and compare results wherever possible. In all cases, the comparison will be between the neutrino mass bounds, as none of the references report constraints on the non-relativistic neutrino energy density.

We start with Ref.~\cite{Cuoco:2005qr}, in which the authors considered a Fermi-Dirac distribution for neutrinos together with a non-thermal component in the form of a Gaussian. 
This reference used a combination of CMB (WMAP, VSA, CBI and ACBAR), large-scale structure data (2dFGRS, SDSS) and type-Ia supernovae data. After performing a standard Bayesian inference analysis, they found that the neutrino mass bound would be relaxed up to $\sum m_\nu \lesssim 1.5\,\mathrm{eV}$ at 95\% CL. This bound should be compared with the one obtained within $\Lambda$CDM using the same data set, $\sum m_\nu \lesssim 0.7\,\mathrm{eV}$~\cite{Cuoco:2005qr}. We believe that following a similar approach with current data, one would find a neutrino mass bound that is fairly similar to $\sum m_\nu \lesssim 0.12\,\text{eV}$, because Planck data roughly requires $N_{\rm eff} \simeq 3 \pm 0.3$ and in their parameterisation they included a Fermi-Dirac component for neutrinos with the temperature expected in the Standard Model (which already gives $N_{\rm eff}^\nu = 3.044$).

In Ref.~\cite{deSalas:2018idd}, the authors considered a neutrino distribution function that interpolates between Fermi-Dirac and Bose-Einstein with the temperature expected in the Standard Model. In this context, the energy density of both ultra-relativistic neutrinos and non-relativistic ones does not differ by more than $30\%$, and consequently the bounds they found for $\sum m_\nu$ are fairly similar to those in $\Lambda$CDM. 

Next, the authors of Ref.~\cite{Oldengott:2019lke} took a more general approach and considered a neutrino distribution function formed by a Fermi-Dirac distribution plus a sum of orthonormal polynomials to the Fermi-Dirac distribution. While in principle any distribution can be decomposed with such polynomials, they restricted their analysis to polynomials up to second order only. Importantly, in each of their main analyses, the prefactors in the expansion were fixed, which means that they did not vary the number density of neutrinos (although it does deviate from the Fermi-Dirac number density in the Standard Model). As a result, using Planck 2015 CMB data and BAO data, they find that the neutrino mass bound is relaxed up to $\sum m_\nu < 0.37\,\mathrm{eV}$ at 95\% CL. This reference does remark that the neutrino mass bound may be relaxed further, depending on the model under consideration. In our work, we have taken a more flexible approach and directly varied the number density of neutrinos $n_{\nu,0}$, in addition to $\sum m_\nu$ and $N_\mathrm{eff}$. This has not only resulted in a robust upper limit on the non-relativistic neutrino energy density today, but also in a significantly looser neutrino mass bound that can be at least as large as the laboratory one $\sum m_\nu \lesssim 3 \, \mathrm{eV}$. 
 
Finally, the authors of Ref.~\cite{Renk:2020hbs} studied the possibility of considering neutrinos following a Fermi-Dirac distribution, but with a smaller temperature than in $\Lambda$CDM. Since $n_\nu \propto T_\nu^3$, this can lead to a substantial relaxation of the neutrino mass bound. However, $T_\nu$ also strongly controls $N_{\rm eff}^\nu \propto T_\nu^4$ and therefore in order to keep consistency with $N_{\rm eff} \simeq 3$, the authors considered a dark-radiation component too. In this context, they showed that the neutrino mass bound can be relaxed up to $\sum m_\nu < 3\,\text{eV}$. Our results are therefore in full agreement with theirs, but we highlight here that they hold for \emph{any} neutrino distribution function. 

\section{BBN and model building}~\label{sec:BBNModelBuilding}

\noindent In this study we have effectively considered the possibility of the neutrino number density being significantly different to the one expected in the standard cosmological model. In practice, we have done this by choosing a suitably flexible neutrino distribution function (see Eq.~\eqref{eq:dist_Gaussian}) for the purpose of studying the consequences on CMB observations. However, in general, this distribution function cannot be the resulting primordial distribution function of neutrinos after Big Bang Nucleosynthesis for two main reasons. Firstly, neutrinos were tightly coupled to the thermal plasma for temperatures $T\gtrsim 2\,\text{MeV}$~\cite{Dolgov:2002wy}, and secondly, the successful prediction of the light element abundances require neutrinos to have a Fermi-Dirac-like distribution at the time at which the proton-to-neutron interactions freeze-out $T\sim 0.7\,\text{MeV}$, see~\cite{Cucurull:1995bx,Dolgov:2005mi,Iizuka:2014wma,deSalas:2018idd}. On the other hand, for temperatures in the range $0.07 \, \mathrm{MeV} \lesssim T \lesssim 0.7\,\text{MeV}$, whilst the precise form of the neutrino distribution function is largely unconstrained -- as neutrinos do not interact with baryons anymore -- there is at least a handle on the energy density of neutrinos, which should be consistent with $N_{\rm eff} \sim 3$. In the context of these physical requirements, we have therefore considered the non-standard neutrino distribution as arising from a mechanism that is only effective at $T \lesssim 0.07\,\text{MeV}$ or, equivalently, $z \lesssim 3\times 10^8$. Furthermore, in our CMB analyses we have fixed the primordial helium abundance to the standard cosmological model prediction $Y_\mathrm{P} = 0.245$~\cite{pdg}, as CMB observations are sensitive to $Y_\mathrm{P}$ too~\cite{Hou:2011ec}.

To summarise, in our setup we consider the neutrino distribution function for $z \lesssim 3 \times 10^8$ to be fixed, but not necessarily thermal. In reality, however, the neutrino distribution should transition from being thermal to non-thermal some time after BBN, and our analysis does not cover scenarios in which such modifications occur arbitrarily late in the expansion history. Essentially, current CMB observations are sensitive to multipoles $\ell$ up to $\ell_{\rm max} \sim \mathcal{O}(10^3)$~\cite{ACT:2020gnv,SPT-3G:2021eoc}. These angular scales probe a wide range of redshifts but are sensitive to the neutrino distribution function only up to a maximum redshift of $z \sim 5\times10^4$, which corresponds to the time at which such perturbation modes enter the horizon. Thus, our results will only strictly apply to scenarios where the neutrino distribution changes in the redshift window\footnote{Note that scenarios with a very low reheating temperature, $T_{\rm RH} \sim 5\,{\rm MeV}$, see e.g.~\cite{deSalas:2015glj,Hasegawa:2019jsa}, or with sterile neutrinos with lifetimes $\tau \sim 0.1\,{\rm s}$~\cite{Boyarsky:2021yoh} could lead to a modification of the neutrino distribution function before BBN. However, in these cases the number density of neutrinos is not too different from the Standard Model expectation, and consequently the neutrino mass bound is not expected to be significantly altered. Similar considerations also apply to distribution functions with non-negligible neutrino chemical potentials, see e.g.~\cite{Steigman:2005uz,Oldengott:2017tzj}.} $5\times10^4 \lesssim z\lesssim 3\times10^8$.

With these clarifications in place, we can now discuss the possible model building opportunities that arise. The cosmological neutrino mass bound within our setup effectively scales as $\sum m_\nu \lesssim 0.12\,\text{eV} \,  n_\nu^{\rm FD} /n_\nu$, which means that in order to substantially relax the cosmological neutrino mass bound, a mechanism allowing for a large reduction of the number density of neutrinos in the early Universe should be invoked. In addition, such a mechanism should ensure that the contribution of neutrinos and dark radiation in the plasma is compatible with $N_{\rm eff} = N_\mathrm{eff}^\nu + N_\mathrm{eff}^\mathrm{DR} \simeq 3.0 \pm 0.4$. As discussed above, this must happen somewhere between $5\times 10^4 \lesssim z\lesssim 3\times 10^8$, in order not to spoil BBN or CMB observations. In this context, there is already one such mechanism in the literature capable of fulfilling this, see~\cite{Farzan:2015pca}. The essence of this idea is to consider massless dark radiation that interacts with a keV scale boson. This boson then subsequently couples to neutrinos and between BBN and recombination can dilute the number density of neutrinos by effectively exchanging them with massless dark radiation. Another interesting alternative is to consider a scenario without any dark radiation. For this to be viable, one would have to reduce the number density of neutrinos whilst at the same time enhancing their mean momentum since $N_{\rm eff}^\nu \propto \bar{p}_\nu n_\nu $ and $N_{\rm eff} \sim 3$. Substantially relaxing the neutrino mass bound in this way would require this process to proceed non-thermally, because $\bar{p} \gg 3\,T_\nu$. Although building such a mechanism is beyond the scope of the current paper, we believe it would be interesting to explore it in light of the consequent relaxation of the neutrino mass bound that could be achieved.

\section{Summary}
~\label{sec:Summary}

\noindent In this work, we have investigated the impact of a non-standard neutrino distribution function on Cosmic Microwave Background measurements. In Sec.~\ref{sec:CMBpheno}, we have argued that the relativistic and non-relativistic neutrino energy densities are the key neutrino properties that one could expect to constrain with CMB data. In Sec.~\ref{sec:ExDists}, to explicitly illustrate this point, we chose five drastically different neutrino distribution functions that nonetheless shared the same non-relativistic and relativistic energy densities, and showed that their impact on the CMB was indistinguishable given the precision of current and future CMB data (see Figs.~\ref{fig:distributions} and~\ref{fig:cmbs4}). In Sec.~\ref{sec:FullAnalysis}, we then took a suitably general parameterisation for the neutrino distribution function which effectively allowed us to independently vary the neutrino mass, number density, and relativistic energy density. Using Planck legacy data, we derived bounds on the effective number of relativistic species $N_\mathrm{eff}$ and non-relativistic neutrino energy density $\rho_{\nu, 0}^\mathrm{NR}$ that are marginalised over the neutrino distribution function (see Fig.~\ref{fig:triangle}). Finally, in Sec.~\ref{sec:BBNModelBuilding}, we discussed the implications of successful BBN on the setup and the relation to the model building of non-standard neutrino distribution functions. The main findings of this work are the following:

\begin{itemize}
    \item \emph{CMB Sensitivity --} Current and future CMB observations are almost completely insensitive to particular features of the neutrino distribution function. Instead, CMB experiments are only capable of meaningfully constraining the relativistic and non-relativistic neutrino energy densities. This further establishes that for non-standard neutrino distributions, whose number density may differ from that of a Fermi-Dirac distribution, CMB observations cannot directly constrain the neutrino mass. 

    \item \emph{Bounds --} We have considered a very flexible neutrino distribution function, that allowed us to independently vary the neutrino number density, relativistic energy density and mass. Given this setting and using Planck 2018 TTTEEE+lowE and BAO data, we obtained bounds on the relativistic and non-relativistic neutrino energy densities of $N_\mathrm{eff} = 3.0 \pm 0.4$ and $\rho_{\nu, 0}^{\mathrm{NR}} < 14 \, \mathrm{eV} \, \mathrm{cm}^{-3}$, respectively, at 95\% CL.
    
    \item \emph{Maximum neutrino number density --} Neutrino oscillation experiments bound the sum of neutrino masses from below to be $\sum m_\nu \geq 0.058 \, \mathrm{eV}$. Thus, given our bound on $\rho_{\nu, 0}^\mathrm{NR}$, we can directly set a mass- and distribution-independent upper bound on the number density of a single neutrino species in the cosmic neutrino background of $n_{\nu, 0} < 241 \, \mathrm{cm}^{-3}$.

    \item \emph{Neutrino mass bound --} By considering the neutrino number density as an independent parameter, we find that the neutrino mass bound is relaxed to at least the end of our prior range -- and to the level of current laboratory limits -- $\sum m_\nu \lesssim 3 \, \mathrm{eV}$. This should be compared to the standard case in which neutrinos follow a Fermi-Dirac distribution, and hence have a fixed number density, where the bound is $\sum m_\nu < 0.12\,\mathrm{eV}$ at 95\% CL~\cite{Akrami:2018vks}. More specifically, the neutrino mass bound scales like $\sum m_\nu \lesssim 0.12\,\text{eV} \,  n_\nu^{\rm FD} /n_\nu$.
    
\end{itemize}

\noindent To summarise, the Cosmic Microwave Background is a powerful probe of the energy density of neutrinos when they are ultra-relativistic and when they are non-relativistic. The neutrino distribution function precisely controls these two quantities and is fixed to be Fermi-Dirac in standard analyses. In this paper, we have demonstrated the full extent to which the CMB is sensitive to modifications to this assumption. In the next section, we will close by discussing the phenomenological consequences of the results presented above.

\section{Outlook}
\label{sec:Outlook}
\noindent The conclusions in the previous section provide us with some interesting perspectives on current and future experimental (neutrino) programs, as well as on particle-physics model building. Here, we will consider some possible future scenarios that could arise, and discuss their interpretation in terms of neutrino properties. There are a number of experiments that are improving or will improve our current understanding of neutrinos. Those relevant in the context of our study can be divided into three categories: \emph{i)} neutrino mass measurements (e.g. the KATRIN~\cite{KATRIN:2001ttj} or Project 8~\cite{Project8:2017nal} experiments, see also~\cite{Formaggio:2021nfz} for a review), \emph{ii)} experiments that could directly detect the Cosmic Neutrino Background (CNB) (e.g. the proposed PTOLEMY~\cite{Baracchini:2018wwj, Betti:2019ouf} experiment), and \emph{iii)} galaxy surveys and other complementary probes of the matter power spectrum (e.g. DESI~\cite{DESI:2016fyo} and EUCLID~\cite{Amendola:2016saw}). A full exploration of the interplay between these probes is beyond the scope of this outlook, however, we discuss exactly this aspect in our recent paper~\cite{Alvey:2021xmq}. It should be noted that PTOLEMY is currently in the research and development phase and is therefore still deciding on a final design specification. On the other hand, experiments such as KATRIN or DESI/EUCLID are already either currently taking data or about to begin their surveys.

Direct measurements of the absolute scale of the neutrino mass are perhaps the easiest to understand in terms of the potential for seemingly inconsistent experimental results. For example, imagine a scenario where the KATRIN experiment --- which is expected to reach a sensitivity of $m_\nu \sim 0.2\, \mathrm{eV}$ by the end of its planned operational period --- makes a positive neutrino mass detection. This would seem to be inconsistent with the cosmological limit derived within the $\Lambda$CDM framework, $\sum m_\nu < 0.12\, \mathrm{eV}$. In the more general context highlighted throughout this paper, however, we can immediately see that a KATRIN detection and a CMB-based bound can be in complete agreement provided that the neutrinos have a number density that is lower than that in $\Lambda$CDM. This plausible experimental scenario would have important implications for neutrinos in cosmology, see~\cite{Alvey:2021xmq}. 

The phenomenology of the latter two experimental categories requires more work to understand the full implications for neutrino properties in cosmology. While the existence of the cosmic neutrino background is indirectly confirmed through observations of the CMB and BBN abundance measurements, its direct detection is an extremely challenging measurement to make, mainly due to the very low characteristic energy of CNB neutrinos and their small interaction cross-section within the SM. One of the currently existing proposals for a methodology of detecting the CNB comes from the PTOLEMY experiment, which looks for the capture of CNB neutrinos on beta-decaying nuclei such as tritium. In the context of this paper, the most important observation regarding this type of experiment is that the sensitivity and detection prospects depend crucially on the neutrino mass and local number density of neutrinos\footnote{This is a combination of the cosmological number density and the clustering enhancement induced by the gravitational pull of the Milky-Way halo, see e.g.~\cite{Ringwald:2004np}.}, which affect the separation from the large beta-decay background and the total rate respectively. The neutrino distribution function directly controls the latter of these, and so there is an intimate link with our current study. As such, we could imagine a PTOLEMY-like experiment being used as a sort of model discriminator, in conjunction with facilities such as KATRIN and DESI/EUCLID, looking for cosmologically sound scenarios with large neutrino masses and/or non-standard neutrino distribution functions. This is exactly the sort of exciting experimental scenario we develop and explore in our recent work~\cite{Alvey:2021xmq}, where we show that there are improved detection prospects at a PTOLEMY-style experiment for large neutrino mass cosmologies, fully consistent with all current cosmological data.

\begin{figure}[t!]
    \centering
    \includegraphics[width=\linewidth]{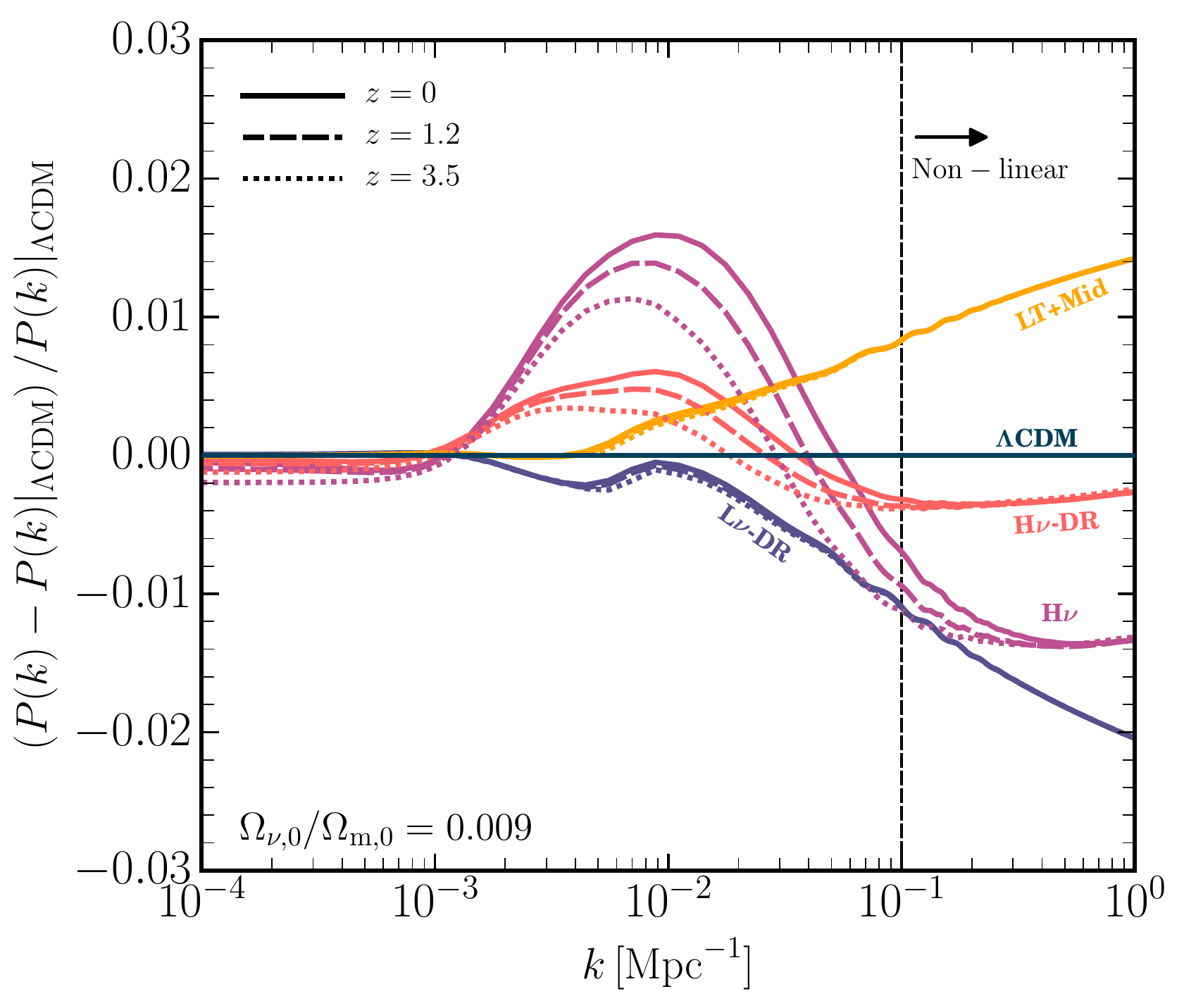}
    \caption{Impact of a non-standard neutrino distribution (detailed in Sec.~\ref{sec:ExDists}) on the linear matter power spectrum compared to the $\Lambda$CDM case with massive neutrinos. The dashed, vertical line highlights at which scales non-linear corrections are relevant. Inferences of the matter power spectrum might provide for an alternative way to differentiate between neutrino distribution functions.}
    \label{fig:mpk}
\end{figure}

Finally, we briefly comment on using the matter power spectrum as a tool to search for features in the neutrino distribution function. Current and upcoming experiments, such as DESI and EUCLID, are aiming to directly measure the sum of neutrino masses, typically with an estimated sensitivity of $\sigma(m_\nu) \sim 0.02 \, \mathrm{eV}$~\cite{DESI:2016fyo, Amendola:2016saw,Font-Ribera:2013rwa, Audren:2012vy}. Given our experience with the CMB, however, we might expect that this may not necessarily be a measurement of the neutrino mass directly, but instead the non-relativistic energy density in neutrinos $\rho_\nu^\mathrm{NR}$. The situation is more complicated than in the CMB case, however, because the effect of neutrinos on the clustering arises as a result of their non-zero free-streaming length and is more pronounced at small scales, where non-linear corrections become relevant. Nevertheless, it could be that even at the linear level ($k \lesssim 0.1\,\mathrm{Mpc}^{-1}$) there already exist some observational hints. We show this in Fig.~\ref{fig:mpk} for the five distribution cases considered in Sec.~\ref{sec:ExDists}, which by construction have the same non-relativistic energy density. We can see that there are clear differences in the matter power spectrum for scales $k \gtrsim 10^{-3} \, \mathrm{Mpc}^{-1}$ in each case. The question is then whether such changes are detectable, and if so, what could we learn about the mass and distribution of neutrinos in the early Universe? The answer to this question would require a careful consideration of up-to-date matter power spectrum predictions -- particularly at the non-linear level -- and experimental uncertainties, see e.g.~\cite{Munoz:2018ajr,Garny:2020ilv,Chen:2020bdf,Archidiacono:2020dvx,Nishimichi:2020tvu, Xu:2020fyg, Xu:2021rwg} for recent studies along these lines within $\Lambda$CDM. Nevertheless, the prospect of a cosmological determination of features in the neutrino distribution function using this probe is a promising alternative and merits further study.
\vspace*{7pt}

\section*{Acknowledgements}
\noindent We are very grateful to Mathias Garny, Isabel Oldengott, Thomas Schwetz and Sunny Vagnozzi for comments on the manuscript.  JA is supported through the research program ``The Hidden Universe of Weakly Interacting Particles" with project number 680.92.18.03 (NWO Vrije Programma), which is partly financed by the Nederlandse Organisatie voor Wetenschappelijk Onderzoek (Dutch Research Council). ME is supported by a Fellowship of the Alexander von Humboldt Foundation. NS is a recipient of a King's College London NMS Faculty Studentship. We acknowledge the use of the public cosmological codes \texttt{CLASS}~\cite{Blas:2011rf,Lesgourgues:2011re} and \texttt{MontePython}~\cite{Audren:2012wb, Brinckmann:2018cvx}. The simulations in this work were performed on the Rosalind research computing facility at King’s College London.

\clearpage
\bibliography{biblio}

\appendix 

\section{Supplementary Plots for Sec. III}
\label{app:supplementary_results_sec3}
\begin{figure*}[t!]
    \centering
    \includegraphics[width=0.9\linewidth]{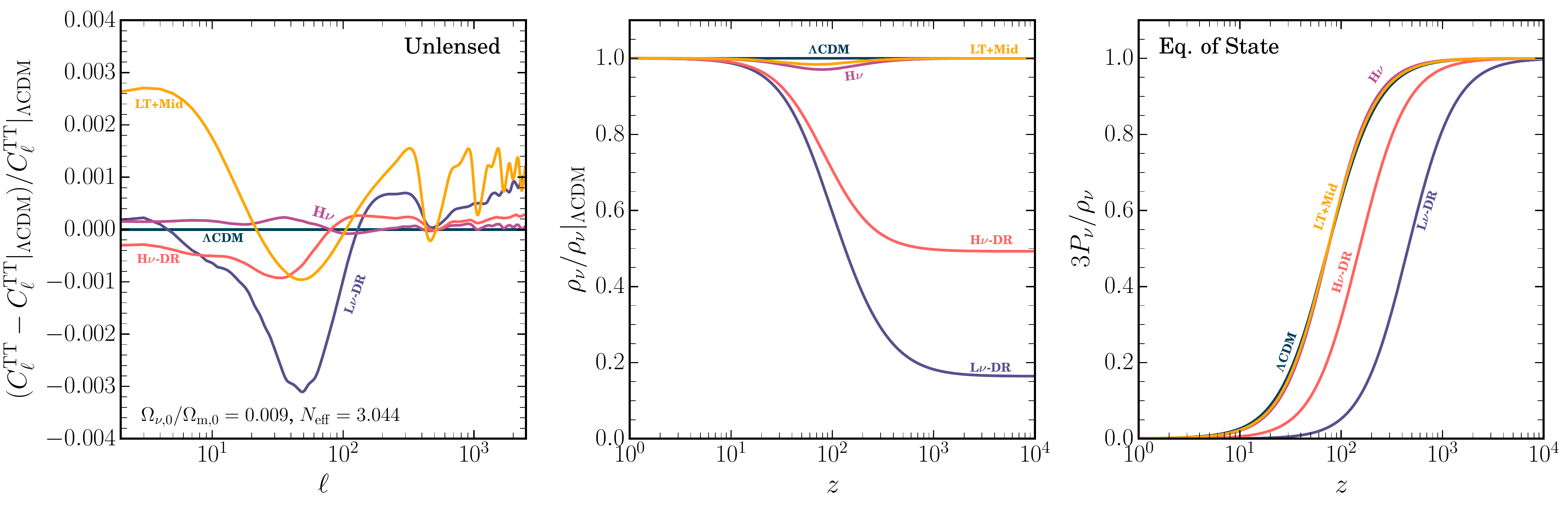}
    \caption{Impact of the neutrino distribution function on the unlensed CMB TT power spectrum (\emph{left}), the neutrino energy density with respect to the SM evolution (\emph{middle}), and the neutrino equation of state (\emph{right}).}
    \label{fig:nu_app}
\end{figure*}

\noindent In this appendix, we show supplementary plots to those displayed in Sec.~\ref{sec:ExDists}. Specifically, in Fig.~\ref{fig:nu_app}, we show how the five discussed distribution functions affect the unlensed CMB TT power spectrum (left panel), the evolution of the neutrino energy density (middle panel), and the neutrino equation of state (right panel).

\section{Full Set of Posteriors}
\label{app:fullsetofposteriors}

\begin{figure*}[h]
    \centering
    \includegraphics[width=0.83\linewidth]{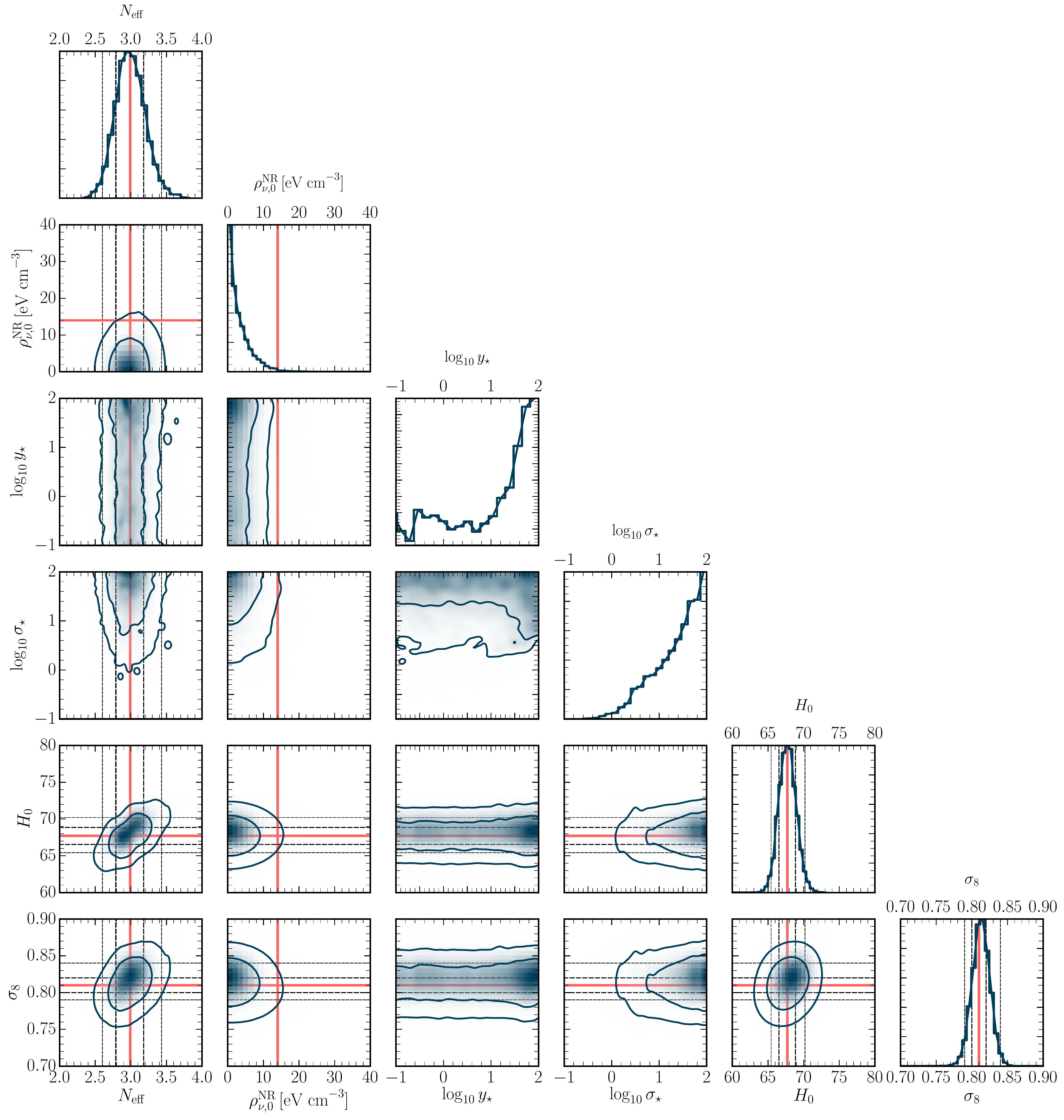}
    \caption{Full set of posteriors from the analysis described in Sec.~\ref{sec:FullAnalysis}, including those for the Gaussian distribution parameters $y_\star$ and $\sigma_\star$, as well as the cosmological parameters $H_0$ and $\sigma_8$.}
    \label{fig:contour_app}
\end{figure*}

In this appendix, we show the full set of posteriors as obtained in our main analysis in Sec.~\ref{sec:FullAnalysis}, see Fig.~\ref{fig:contour_app}. These include: the effective number of relativistic species $N_\mathrm{eff}$, the non-relativistic neutrino energy density $\rho_{\nu,0}^\mathrm{NR}$, the average $y_*$ and width $\sigma_*$ of the Gaussian distribution in Eq.~\eqref{eq:dist_Gaussian}, the Hubble constant $H_0$, and $\sigma_8$.

\section{The Transition from Radiation to Matter}\label{app:Transition}

\noindent An important characteristic of massive neutrinos in cosmology as compared to their massless counterparts is that at some point in the Universe's evolution, they transition from being a radiation component of the Universe, with $P_\nu / \rho_\nu \sim 1/3$, to a pressureless matter component. In the main text, we estimated the redshift at which this occurred, but did not show the full derivation. Here, we outline the complete computation.

To begin, we note that the energy density of relativistic neutrinos is given by $\rho_\nu(z \gg z_\mathrm{NR}) = \bar{p}(z) n_\nu(z)$. So we can compute the average momentum $\bar{p}(z \gg z_\mathrm{NR})$ directly in terms of $\rho_\nu$, which is proportional to $N_\mathrm{eff}$ at this time, and $n_\nu(z) = n_{\nu,0} (1 + z)^3$, where $n_{\nu,0}$ is the neutrino number density today. Using the fact that we can write
\begin{align}
    \rho_\nu(z \gg z_\mathrm{NR}) = \frac{7}{8}\left(\frac{4}{11}\right)^{4/3} N_\mathrm{eff}^\nu \rho_\gamma(z \gg z_\mathrm{NR}) \nonumber \\ = \frac{7}{8}\left(\frac{4}{11}\right)^{4/3} N_\mathrm{eff}^\nu \rho_{\gamma, 0}(1 + z)^4\ ,
\end{align}
then we can compute the average momentum at some high redshift via:
\begin{align}
    \bar{p}(z \gg z_\mathrm{NR}) &= \frac{7}{8}\left(\frac{4}{11}\right)^{4/3} \frac{N_\mathrm{eff}^\nu \rho_{\gamma, 0}(1 + z)^4}{n_\nu(z \gg z_\mathrm{NR})} \nonumber \\ &= \frac{7}{8}\left(\frac{4}{11}\right)^{4/3}\frac{N_\mathrm{eff}^\nu \rho_{\gamma, 0}(1 + z)}{n_{\nu, 0}} \ .
\end{align}
If we use the redshift relation that $\bar{p}(0) (1 + z) = \bar{p}(z)$, now for any redshift, then we can obtain the average momentum of the neutrinos today as:
\begin{equation}
    \bar{p}(0) = \frac{\bar{p}(z \gg z_{\mathrm{NR}})}{1 + z} = \frac{7}{8}\left(\frac{4}{11}\right)^{4/3} \frac{N_\mathrm{eff}^\nu}{n_{\nu, 0}} \rho_{\gamma, 0}\ .
\end{equation}
We can then directly compute an estimate for the redshift $z_\mathrm{NR}$ by comparing the average momentum of the particles to the mass of the neutrino $\bar{p}(z_{\mathrm{NR}}) \sim m_\nu \sim \frac{1}{3}\sum m_\nu$. Then, since $\bar{p}(z_\mathrm{NR}) = \bar{p}(0) (1 + z_\mathrm{NR})$, we find:
\begin{align}
    1 + z_\mathrm{NR} \sim \frac{\frac{1}{3}\sum m_\nu}{\bar{p}(0)} = \frac{8}{7}\left(\frac{11}{4}\right)^{4/3} \frac{1}{3} \left(\frac{\sum m_\nu n_{\nu, 0}}{\rho_{\gamma, 0}N_\mathrm{eff}^\nu}\right)\ .
\end{align}
If we identify the non-relativistic energy density today as $\rho_{\nu, 0}^{\mathrm{NR}} = \sum m_\nu n_{\nu, 0}$, then we obtain the result given in Sec.~\ref{sec:ExDists} that the redshift at which neutrinos become non-relativistic depends only the ratio between $\rho_{\nu, 0}^{\mathrm{NR}}$ and $N_\mathrm{eff}^\nu$:
\begin{align}
    1 + z_\mathrm{NR} &= \frac{8}{7}\left(\frac{11}{4}\right)^{4/3} \frac{1}{3\rho_{\gamma, 0}} \left(\frac{\rho_{\nu, 0}^{\mathrm{NR}}}{N_\mathrm{eff}^\nu}\right) \nonumber \\ &\sim 56 \, \frac{\rho_{\nu, 0}^{\mathrm{NR}}}{30 \, \mathrm{eV} \, \mathrm{cm}^{-3}} \, \frac{3.044}{N^\nu_{\mathrm{eff}}}\ .
\end{align}

\newpage

\end{document}